\documentclass[]{emulateapj}

\usepackage{amssymb,natbib,graphicx,subfigure,apjfonts,color}

\def\kms{{\rm\thinspace km s}^{-1}}

\def\pcmsq{{\rm\thinspace cm}^{-2}}
\def\pcmcb{{\rm\thinspace cm}^{-3}}

\def\kpc{{\rm\thinspace kpc}}

\def\Mpc{{\rm\thinspace Mpc}}   
\def\Msun{\hbox{$\rm\thinspace M_{\odot}$}}
\def\pc{{\rm\thinspace pc}}     
       
\def\yr{{\rm\thinspace yr}}
\def\Myr{{\rm\thinspace Myr}}
\def\Gyr{{\rm\thinspace Gyr}}
\def\Msunyr{\Msun\yr^{-1}}
\def\Msunpc2{{\Msun\pc}^{-2}}
\def\Msunyrkpc2{{\Msun\yr^{-1}\kpc}^{-2}}
\def\Ha{{\rm H\alpha}}
\def\magarcsec2{{\rm\thinspace mag\thinspace arcsec}^{-2}}
\def\ergs{{\rm\thinspace erg s}^{-1}}

\shorttitle{Short-lived clumps in simulated $z\approx2$ disks}
\shortauthors{Genel et al.}

\begin{document}

\title{Short-lived star-forming giant clumps in cosmological simulations of $\lowercase{z}\approx2$ disks}

\author{Shy Genel\altaffilmark{1,2}, Thorsten Naab\altaffilmark{3}, Reinhard Genzel\altaffilmark{1,4}, Natascha M. F{\"o}rster Schreiber\altaffilmark{1}, Amiel Sternberg\altaffilmark{2}, Ludwig Oser\altaffilmark{3,5}, Peter H. Johansson\altaffilmark{6,7}, Romeel Dav{\'e}\altaffilmark{8}, Benjamin D. Oppenheimer\altaffilmark{9}, Andreas Burkert\altaffilmark{5}}

\altaffiltext{1}{Max Planck Institut f\"ur extraterrestrische Physik, Giessenbachstrasse, 85748 Garching, Germany; shy@mpe.mpg.de; genzel@mpe.mpg.de; forster@mpe.mpg.de}
\altaffiltext{2}{Sackler School of Physics and Astronomy, Tel Aviv University, Tel Aviv 69978, Israel; amiel@wise.tau.ac.il}
\altaffiltext{3}{Max Planck Institut f\"ur Astrophysik, Karl-Schwarzschild-Str. 1, 85741 Garching, Germany; naab@mpa-garching.mpg.de; oser@usm.lmu.de}
\altaffiltext{4}{Department of Physics, Le Conte Hall, University of California, Berkeley, CA 94720}
\altaffiltext{5}{Universit\"ats-Sternwarte M\"unchen, Scheinerstr.\ 1, D-81679 M\"unchen, Germany; burkert@usm.lmu.de}
\altaffiltext{6}{Department of Physics, University of Helsinki, Gustaf H\"allstr\"omin katu 2a, FI-00014 Helsinki, Finland}
\altaffiltext{7}{Finnish Centre for Astronomy with ESO, University of Turku, V\"ais\"al\"antie 20, FI-21500 Piikki\"o, Finland; phjohans@astro.helsinki.fi}
\altaffiltext{8}{Astronomy department, University of Arizona, Tuscon, AZ 85721; rad@as.arizona.edu}
\altaffiltext{9}{Leiden Observatory, Leiden University, PO Box 9513, 2300 RA Leiden, the Netherlands; oppenheimer@strw.leidenuniv.nl}

\slugcomment{The Astrophysical Journal, accepted}

\begin{abstract}
Many observed massive star-forming $z\approx2$ galaxies are large disks that exhibit irregular morphologies, with $\approx1\kpc$, $\approx10^{8-10}\Msun$ clumps. We present the largest sample to date of high-resolution cosmological SPH simulations that zoom-in on the formation of individual $M_*\approx10^{10.5}\Msun$ galaxies in $\approx10^{12}\Msun$ halos at $z\approx2$. Our code includes strong stellar feedback parameterized as momentum-driven galactic winds. This model reproduces many characteristic features of this observed class of galaxies, such as their clumpy morphologies, smooth and monotonic velocity gradients, high gas fractions ($f_{\rm g}\approx50\%$) and high specific star-formation rates ($\gtrsim1\Gyr^{-1}$). In accord with recent models, giant clumps ($M_{\rm clump}\approx(5\times10^8-10^9)\Msun$) form in-situ via gravitational instabilities. However, the galactic winds are critical for their subsequent evolution. The giant clumps we obtain are short-lived and are disrupted by wind-driven mass loss. They do not virialise or migrate to the galaxy centers as suggested in recent work neglecting strong winds. By phenomenologically implementing the winds that are observed from high-redshift galaxies and in particular from individual clumps, our simulations reproduce well new observational constraints on clump kinematics and clump ages. In particular, the observation that older clumps appear closer to their galaxy centers is reproduced in our simulations, as a result of inside-out formation of the disks rather than inward clump migration.
\end{abstract}


\section{Introduction}
\label{s:intro}
Star-forming disk galaxies at $z\approx2$ differ from their local counterparts in several aspects: they are more gas-rich and rapidly star-forming, and have irregular morphologies and high gas velocity dispersions. Spatially resolved observations of their rest-frame UV and optical continuum light often reveal large ($\approx1\kpc$) clumps \citep{ElmegreenB_05b,ElmegreenB_09a,FoersterSchreiberN_11b}. Clumpy morphologies are also observed with star-formation tracers such as $\Ha$ (e.g.~the SINS survey; \citealp{GenzelR_08a,GenzelR_11a}). The spectrally resolved line emissions reveal ordered rotation and high velocity dispersions, widespread across the disks at $\approx50-80\kms$ \citep{GenzelR_06a,GenzelR_08a,FoersterSchreiberN_06a,FoersterSchreiberN_09a,ShapiroK_08a,CresciG_09a,LawD_09a,JonesT_10a,SwinbankM_10a}.

Theoretically, the large $z\approx2$ star-forming galaxies have been considered as marginally-unstable thick disks, where the large random motions set the large masses and sizes of unstable regions that collapse into giant clumps \citep{ImmeliA_04a,BournaudF_07a,BournaudF_08a,GenzelR_08a,ElmegreenB_08a,AgertzO_09a,DekelA_09b,BurkertA_10a,CeverinoD_09a}. In this picture, the high star-formation rates result from the high growth rates of cosmic structures at $z\approx2$ \citep{GenelS_08a}, as gas funnels directly from the cosmic web to the vicinity of galaxies in a 'cold-mode' accretion \citep{KeresD_05a,KeresD_09a,DekelA_09a}. However, several observational clues have not yet been addressed in this framework. First, vigorous outflows are observed from a variety of star-forming galaxies at high redshift, both globally (e.g.~\citealp{PettiniM_00a,WeinerB_09a,ShapiroK_09a,SteidelC_10a}), and from individual giant clumps \citep{GenzelR_11a}. Second, high signal-to-noise observations of individual giant clumps reveal only minor kinematical signatures, sometimes indicating dynamical masses lower than alternative independent clump mass estimates \citep{GenzelR_11a}. Third, comparisons of abundance and clustering of galaxies and dark matter halos suggests that at $z\approx2$, much like at $z=0$, typically not more than $\approx10\%$ of the baryons associated with halos of any mass have turned into stars \citep{MosterB_09a,BehrooziP_10a,WakeD_11a}, causing some tension with the high efficiency of 'cold-mode' accretion.

Further on, one aspect of the theoretical picture that developed in recent years is that the giant clumps lose angular momentum and energy due to dynamical friction in the disk, clump-clump interactions, and internal torques within the disk, and thereby spiral in to the center of the galaxy. This has been based on two main lines of argument. First, age gradients have been inferred in clumpy disks from the presence of a redder/older component often coinciding or close to the galaxy center \citep{vanderBerghS_96a,ElmegreenB_09a}. Also, a correlation has been found between the relative dynamical dominance of that central component to a (metallicity-based) age estimate for the system, which has been interpreted as indicative of a time sequence, where older systems have had more time to transform part of their disks into bulges via clump migration \citep{GenzelR_08a}. Second, numerical simulations, both of isolated unstable disks \citep{ImmeliA_04b,BournaudF_07a,ElmegreenB_08a,BournaudF_09a} and of cosmological $z\approx2$ disks \citep{AgertzO_09a,CeverinoD_09a}, find the clumps to evolve in that fashion, and theoretical considerations also suggest there should be a significant inflow rate inside the disk \citep{DekelA_09b,BournaudF_11a}. Recently \citet{FoersterSchreiberN_11b} used deep higher-resolution NIC2 imaging to identify more robustly clumps in rest-frame optical emission, and, in combination with AO-assisted SINFONI $\Ha$ observations and ACS imaging, radial trends in clump stellar age with distance to galaxy center were inferred. This result appears consistent with the theoretical inspiraling picture, as noted by \citet{CeverinoD_11a}.

Therefore, a crucial issue is whether giant clumps can survive long enough to become stellar-dominated and/or sink to the centers of their galaxies, contributing to the buildup of bulges. Conflicting theoretical claims have been made \citep{MurrayN_10a,KrumholzM_10a}, while probing that question directly with observations is a difficult task. Therefore, the qualitative agreement between simulations and observations has been interpreted as supporting the picture of long-lived migrating clumps. However, none of the cosmological or isolated simulations reported so far (but see \citealp{SalesL_10a}), which find long-lived migrating clumps, include strong galactic winds, which as noted already are directly observed. As the 'clumpy phase' appears to be ubiquitous at $z\approx1-3$, it is of fundamental importance for our understanding of galaxy formation to know what role giant clumps play in building bulges/spheroids.

In this paper, we demonstrate as a proof of principle that incorporating galactic superwinds resulting from stellar feedback into the existing theoretical picture, can have important implications for our understanding of $z\approx2$ disks and their evolution. We use hydrodynamical cosmological simulations to investigate the formation and properties of a sample of star-forming disks. We use a phenomenological model for the generation of galactic winds to test a scenario in which galactic superwinds are a major driving force in shaping galaxy disks at high redshift. The phenomenological a priori parameters of our model match the observed properties of winds in the type of galaxies we are considering. Our results are heavily dependent on this phenomenology, hence the uncertainty in the observations of such winds is a source of uncertainty for the basic assumptions of our model. However, we find that the implications of implementing such winds in simulations, namely that the giant clumps are short-lived transient features, and thus do not survive to migrate to the centers of their galaxies to form a bulge, are consistent with the existing observational constraints. Hence, we propose a new scenario for the evolution of $z\approx2$ clumpy disks.

\section{The Simulations: Code and Setup}
\label{s:simulations}
We run 'zoom-in' cosmological simulations focused on individual halos taken from the $72h^{-1}\Mpc$ cosmological dark matter simulation presented in \citet{OserL_10a} for the following cosmological parameters: $\Omega_m=0.26$, $\Omega_{\Lambda}=0.74$, $\Omega_b=0.044$, $h=0.72$, $n=0.95$ and $\sigma_8=0.77$. As we focus on massive disks with high star-formation rates (SFR), we select at $z=2$ halos of $\approx10^{12}\Msun$ with instantaneous dark matter growth rates exceeding $500\Msunyr$ and that had no major merger (mass ratio $<3:1$) since $z=3$ (see \citealp{GenelS_08a}). The second criterion selects approximately half of the halos of that mass at that redshift, and together these criteria are met by $\approx15\%$ of those halos, i.e.~our galaxies form in halos that are neither the most 'typical' nor very 'special' or rare. From this sample, we randomly selected nine halos and generated 'zoom-in' initial conditions for re-simulations including baryons, as in \citealp{OserL_10a,OserL_11a}. Our scheme results in high-resolution regions of $\approx5\Mpc$ comoving, on a side, i.e.~approximately $1000$ times larger (in volume) than our main halos at $z\approx2$. The mass resolution of our simulations is $8\times10^5\Msun$ and $5\times10^6\Msun$ for baryonic and high-resolution dark matter particles, respectively. The gravitational softening lengths of baryonic particles are $200h^{-1}\pc$, constant in comoving units, resulting in physical softening lengths of $\approx90\pc$ at $z=2$. 

The cosmological box is evolved to $z=2$ using a version of the N-body/SPH code {\it Gadget-2} \citep{SpringelV_05c} that is very similar to the one described in \citet{OppenheimerB_06a,OppenheimerB_08a}. This version includes ionisation and heating by a uniform background radiation \citep{HaardtF_01a} in the optically thin limit, atomic cooling down to $T=10^4K$ from hydrogen and metals, star-formation and feedback, as well as mass loss and metal enrichment from AGB stars and supernovae of types II and Ia.

The \citet{OppenheimerB_08a} version of the code uses the \citet{SpringelV_03a} star-formation prescription, where stars form stochastically from the gas according to a \citet{SchmidtM_59a} law that reproduces the \citet{KennicuttR_98a} gas surface density to star-formation relation. In addition, that prescription includes a sub-grid model for supernova feedback that results in an effective equation of state for the star-forming gas $P\propto\rho^{\gamma(\rho)}$, where $P$ is the gas pressure, and $\rho$ its physical density. The running index $\gamma(\rho)$ in the \citet{SpringelV_03a} model is rather high at low and medium densities (a stiff equation of state), but becomes negative at high densities, such that the thermal Jeans mass $M_J$ can rapidly drop below the resolution of the simulation \citep{HopkinsP_10b}. The stiffness of this equation of state at medium densities allows construction of stable disk models (e.g.~\citealp{SpringelV_05d}). It is less relevant for simulations of marginally-unstable disks. Therefore, we replaced the star-formation model in the \citet{OppenheimerB_08a} code with the \citet{SchayeJ_08a} approach. There, the usual cooling and heating operate at $\rho<\rho_{EOS}$, while for $\rho>\rho_{EOS}$ a polytropic equation of state with $\gamma=4/3$ is implemented. This ensures that the (thermal) Jeans mass $M_J$ remains constant with varying density and does not become unresolved. Here $\rho_{EOS}=0.1\pcmcb$ is defined such that $M_J\approx1.4\times10^7\Msun$ at $\rho=\rho_{EOS}$ and $T=10^4K$. Star-formation is parameterised as
\begin{eqnarray}
\frac{d\rho_*}{dt}=\frac{\rho}{1\Gyr}(\frac{\rho}{\rho_{th}})^{0.26},
\label{e:sf_law}
\end{eqnarray}
with the star-formation density threshold $\rho_{th}=\rho_{EOS}=0.1\pcmcb$. This reproduces the \citet{KennicuttR_98a} star-formation relation \citep{SchayeJ_08a}. A further addition we introduced to the code is that we calculate the local minimum (over all directions) of the gas surface density $\rho^2/|\nabla\rho|$ \citep{GnedinN_09a}, and suppress star-formation if $\rho^2/|\nabla\rho|<10^{21}\pcmsq$. This modification has only minor consequences for our results.

The feedback scheme developed by \citet{OppenheimerB_06a,OppenheimerB_08a} builds on the kinetic feedback scheme of \citet{SpringelV_03a}, where star-forming gas particles are stochastically kicked with velocities in the ${\bf v\times a}$ direction, where ${\bf v}$ is the particle's velocity prior to the kick and ${\bf a}$ its acceleration. \citet{OppenheimerB_06a,OppenheimerB_08a} introduced a significant modification to this mechanism by tuning the two parameters that control the wind. In their prescription, the magnitude of the kick is $v_{wind}=\sigma(2+3\sqrt{f_L-1})$ and the mass-loading factor
\begin{eqnarray}
\frac{\dot{M}_W}{SFR}\equiv\eta=\frac{\sigma_0}{\sigma},
\label{e:mass_loading}
\end{eqnarray}
where $f_L\approx1-2$ is the luminosity factor, $\sigma$ is the 'velocity dispersion' of the galaxy that is calculated from its mass \footnote{Galaxy dynamical masses are calculated based on an on-the-fly group finder, and then used to calculate $\sigma\equiv200[(M_{\rm dyn}/5\cdot10^{12}\Msun)h(H(z)/H_0)]^{1/3}$. For calculation efficiency, a friends-of-friends (FOF) algorithm is used. We use a linking length $b=0.025(\frac{H(z)}{H(0)})^{1/3}$ that is an improvement over the constant $b$ used in \citet{OppenheimerB_08a}, since it is calibrated to the SKID group finder at various redshifts.}, and $\sigma_0$ is a constant (here $\sigma_0=300\kms$, as in \citet{OppenheimerB_06a}). Introducing $\sigma$-dependencies, these two parameters then scale with the mass of the galaxy, following the theory of momentum-driven winds induced by the radiation pressure from young stars \citep{MurrayN_05a,ZhangD_10a}.

The momentum budget available from feedback from young stars is somewhat uncertain. \citet{MurrayN_05a} estimate the momentum injection from radiation pressure of OB stars (per unit stellar mass formed) to be $\approx200\kms$, and a similar amount of momentum from supernova explosions. \citet{OstrikerE_11a}, however, estimate the specific momentum available from supernovae to be $\approx3000\kms$. In addition, high dust column densities can provide large optical depths that increase the amount of momentum injection by radiation pressure by factors of several (e.g.~\citealp{MurrayN_10a}). Finally, stellar winds can also inject similar amounts of momentum \citep{LeithererC_99a}. We use $v_{wind}=\sigma(4+4.29\sqrt{f_L-1})$, such that the value that we use for the momentum injection per unit stellar mass formed, $\eta v_{wind}\lesssim2500\kms$, is a physically plausible value, even if on the high side. Note that we use higher wind velocities than in \citet{OppenheimerB_08a}. This ensures the escape of all wind particles from the disk, in particular from regions in the disk that have higher escape velocities than the mean one. This change is necessary, since we have a significantly higher resolution in this work, such that the center of the galaxy may become dense enough for its escape velocity to be higher than that at the edge of the disk.

The scalings of the momentum-driven winds are in accordance with observational evidence for the scalings of galactic winds (e.g.~\citealp{MartinC_05a,RupkeD_05a}). The typical velocities we obtain with this model for the central galaxies considered are $\approx400-700\kms$, in agreement with observations. The typical mass-loading factor for our central galaxies is $\eta\approx4$, also in order-of-magnitude agreement with observations, although the observational determination of the mass-loading factors is far less certain than that of the wind velocities (e.g.~\citealp{ErbD_06a,GenzelR_11a} and references therein).

In the \citet{SpringelV_03a} and \citet{OppenheimerB_08a} models, subsequent to being given a velocity kick, gas particles are temporarily decoupled from hydrodynamic interactions to allow them to propagate out of their star-formation sites and eventually leave their galaxies. When simply giving gas particles momentum kicks rather than using a more sophisticated physical model (in a better-resolved interstellar medium), the only way to achieve winds that escape the vicinity of the galaxies (rather than efficiently becoming a galactic fountain) is by using the temporary decoupling method \citep{DallaVecchiaC_08a}. Moreover, even high resolution simulations of isolated galaxies have so far had difficulties in forming efficient winds when having full hydrodynamics and cooling (e.g.~\citealp{MacLowM_99a,FujitaA_04a,DuboisY_08a}). However, such winds are observed, and emerge out of $z\approx2$ galaxies to large distances \citep{SteidelC_10a}. Therefore, we keep the decoupling prescriptions adopted by \citet{OppenheimerB_08a}. We have, however, modified the \citet{OppenheimerB_08a} code to ensure that the velocity kicks are perpendicular to the disk (e.g.~\citealp{BordoloiR_11a}, Bouch{\'e} et al.~in prep.). This is done by calculating the direction of the velocity kicks wind particles are given as ${\bf (v-v_{\rm gal})\times(a-a_{\rm gal})}$, where ${\bf v}$, ${\bf a}$ are the velocity and acceleration of the launched wind particle just before the kick, and ${\bf v_{\rm gal}}$, ${\bf a_{\rm gal}}$ are the velocity and acceleration of the center-of-mass of the galaxy, i.e.~${\bf v\times a}$ is calculated in the inertial frame of the galaxy.

The theoretical considerations and actual observations of winds from galaxies are augmented by further motivation for using the \citet{OppenheimerB_08a} model, namely this strong galactic winds model succesfully matches several important observations:
\begin{itemize}
\item
Properties of the intergalactic medium, such as metallicity and ionization state \citep{OppenheimerB_06a,OppenheimerB_09a,DaveR_10a}.
\item
The galaxy mass-metallicity relation \citep{FinlatorK_08a,DaveR_11b}.
\item
High galactic gas fractions at high redshift (Section \ref{s:global_properties} and \citealp{DaveR_11b}).
\item
Low fraction of cosmic baryons in galaxies at all redshifts (Section \ref{s:global_properties} and \citealp{DaveR_11a}).
\end{itemize}
It should be kept in mind, however, that the simulations presented in this work have much higher resolution than previous works that employed the \citet{OppenheimerB_08a} model. Therefore, some adverse side effects cannot be excluded before a large cosmological box is run with a similarly high resolution. Indeed, we corrected for one such side effect by using increased velocities (see above).

\section{Results}
\label{s:disks}
\subsection{Global properties}
\label{s:global_properties}

\begin{figure}[tbp]
\centering
\includegraphics[]{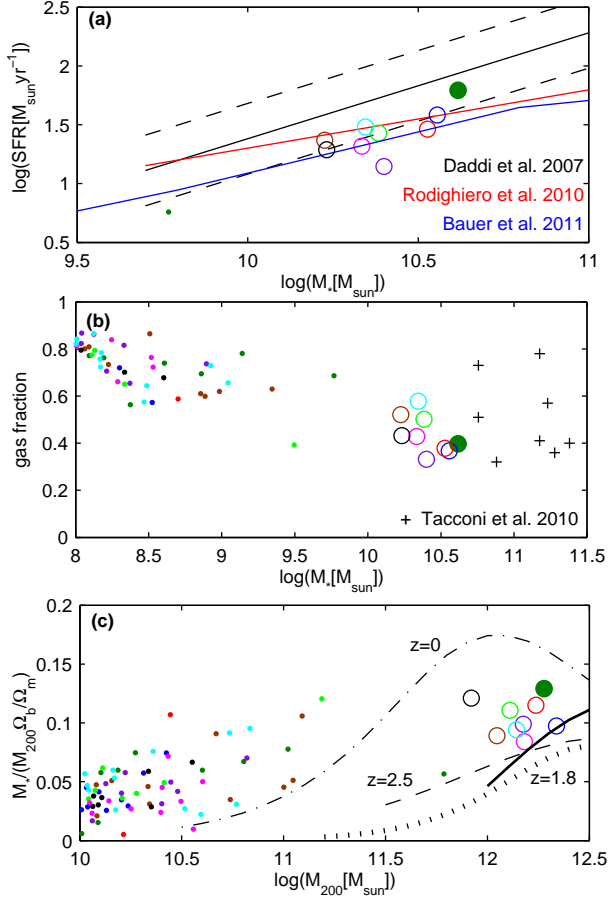}
\caption{Global properties of nine 'zoom-in' simulations at $z=2$, where each color represents one simulation, with its main galaxy ({\it circles}) and the smaller galaxies that formed in neighbouring dark matter halos inside the zoomed-in region ({\it dots}). Only galaxies that are central in their own halos are shown, i.e.~leaving out satellite galaxies that are already inside the main zoomed-in halo. In all panels {\it s224} is marked with a filled circle. {\it Top}: SFR versus stellar mass. The observed \citet{DaddiE_07a} relation for $z=2$ ({\it solid}; standard deviation is {\it dashed}), \citet{RodighieroG_10a} relation for $1.5<z<2$ ({\it red}) and \citet{BauerA_11a} relation for $1.5<z<3$ ({\it blue}), all corrected for a \citet{ChabrierG_03a} initial mass function, are plotted on top of the simulation results, and demonstrate a reasonable match between our simulated galaxies and observations. {\it Middle}: Gas fractions (gas mass over baryonic mass) versus stellar mass. There is a trend of increasing gas fractions with decreasing mass, and the fractions for our main sample are $\approx40\%$, in good agreement with observations at $z\approx2.2$ (\citealp{TacconiL_10a}, {\it pluses}). {\it Bottom}: 'Baryon conversion efficiency' as a function of halo mass. The highest efficiencies of $\approx10\%$ are obtained at $M_{halo}\gtrsim10^{12}\Msun$, in rough agreement with results obtained from matching observations to dark matter halo models, which are taken from \citet{MosterB_09a} for $z=2.5$ ({\it dashed}), $z=1.8$ ({\it dotted}) and $z=0$ ({\it dashed-dotted}), as well as from \citet{WakeD_11a} for $1<z<2$ ({\it solid}). In all panels, the baryonic quantities (i.e.~all but $M_{200}$) are computed inside $3R_{1/2}$, where $R_{1/2}$ is the stellar half-mass radius within $0.2R_{200}$.}
\vspace{0.5cm}
\label{f:global_relations}
\end{figure}

We begin by considering several global properties of the galaxies formed in our sample of nine 'zoom-in' simulations and comparing them with observational constraints. Table \ref{t:properties} gives halo masses, peak circular velocities, sizes, gas and stellar masses, gas fractions and SFRs for the main galaxies in our sample. This sample can be compared to a 'median disk' from the SINS survey, whose properties are given in Table \ref{t:properties} as well. Bearing in mind the uncertainties in the observational quantities, which are of roughly a factor of two for the masses and SFRs, and the unavoidable differences in the definitions of some of the quantities between simulations and observations (see footnotes to Table \ref{t:properties}), it seems that our simulated galaxy sample is overall a good match to the disks observed in the SINS survey. Nevertheless, our simulated sample may be somewhat 'scaled-down' compared to the observed sample, by a factor of approximately two in the masses and SFRs, and less than that in size.

In Figure \ref{f:global_relations}, panel (a) shows the SFR versus stellar mass ($M_*$) plane for central galaxies in the high-resolution region of each of our simulations. The simulated galaxies populate the same region as observed $z\approx2$ galaxies, taken from \citet{DaddiE_07a} ({\it black}), \citet{BauerA_11a} ({\it blue}) and \citet{RodighieroG_10a} ({\it red}). The scatter around the mean \citet{DaddiE_07a} relation is roughly $0.5{\rm dex}$ ({\it dashed}), and the uncertainty in those measurements is of similar magnitude, as evident from the differences between the three lines. The SFR history in the central galaxies is rather constant, as demonstrated in Table \ref{t:properties} by comparing the current SFR to its past average. Figure \ref{f:global_relations}(b) presents gas fractions ($f_{\rm g}\equiv\frac{M_{\rm g}}{M_{\rm g}+M_*}$) versus stellar mass. There is a trend of increasing gas fractions with decreasing mass (see also \citealp{DaveR_11b}), and the fractions for our main sample are $\approx40\%$, in good agreement with observations at $z\approx2.2$ (\citealp{TacconiL_10a}, {\it pluses}). Figure \ref{f:global_relations}(c) shows 'baryon conversion efficiencies' ($\frac{M_*}{M_{200}\Omega_b/\Omega_m}$) versus halo mass. Our simulated halos are presented ({\it symbols}) alongside the same quantity as it is estimated from comparisons of galaxy and halo clustering and abundances (\citealp{MosterB_09a,WakeD_11a}; {\it curves}). The baryon conversion efficiency is a strong function of mass in our simulations, though not as strong as observations indicate, such that small dark matter halos in our simulations probably host too massive galaxies (see also \citealp{DaveR_11a}).

These observed correlations are important constraints that should be reproduced in realistic simulated galaxies. The SFR versus stellar mass `main sequence' relation has received significant attention in recent years and it is a key constraint for galaxy evolution. The gas fraction of galaxies is a dominant factor in their dynamics, and the baryonic conversion efficiency controls the number density of galaxies. Regardless of the imperfect agreement with observations (in particular at halo/galaxy masses below the scale that is the focus of this paper), the global correlations in Figure \ref{f:global_relations} are a result of the galactic winds model we are using, without which they would compare much worse to the observations \citep{DaveR_08a,DaveR_09a,DaveR_11b,DaveR_11a}. For example, the three clumpy disks simulated by \citet{CeverinoD_09a} in a cosmological context with less effective feedback have gas fractions as low as $18\%,14\%,4\%$ and baryon conversion efficiencies as high as $0.26,0.22,1.14$, both indicating too little suppression of star-formation when compared to observations.

It is apparent that our prescriptions dump more momentum per baryon than used in, e.g., \citet{CeverinoD_09a}, \citet{BournaudF_07a} and \citet{AgertzO_09a}. As such, our sub-grid models may seem extreme and the negative impact of winds on the clump survival (Section \ref{s:clump_evolution}) overdone. However, consider what would happen if we switch off this recipe, or substantially turn down its efficacy. To this end, we have performed additional experiments, where we used less efficient winds, either by injecting less momentum (using slower winds, or winds with lower mass-loading factors), or by not temporarily decoupling the wind particles from the hydrodynamics. If the winds are not decoupled they cannot escape the vicinity of the galaxies and are ineffective \citep{DallaVecchiaC_08a}. In these experiments, we obtained galaxies with very different properties, in agreement with similar results from the literature (reviewed above), and in contrarst with observations. For example, when halo {\it s224} is simulated when winds are not temporarily decoupled from the hydrodynamics, we obtain a central galaxy at $z=2$ with low specific star formation rate ($0.6\Gyr^{-1}$ compared to $1.5\Gyr^{-1}$ in the fiducial model), low gas fraction ($8.6\%$ compared to the fiducial $40\%$), high baryonic conversion efficiency ($0.26$ compared to the fiducial $0.129$), high (and peaked) rotation curve (with the peak at $555\kms$ compared to the fiducial $278\kms$), and small size ($R_{1/2}=1\kpc$ compared to $R_{1/2}=3.3\kpc$). Moreover, the clumpy nature of the galaxies, discussed in detail in the following sections, is also lost under such conditions.

\begin{table*}[tbp]
\caption{Galaxy and clump properties}
\label{t:properties}
\centering          
\begin{tabular}{ c | c  c  c  c  c  c  c  c  c  c } 
\hline\hline 
& halo mass & baryon conversion & peak & Half-mass & Stellar mass & Gas mass & Gas fraction & SFR\tablenotemark{(b)} & Number of & ${\rm\frac{SFR}{\langle SFR\rangle}}$\tablenotemark{(d)}\\
& $M_{200}$ & efficiency & $Vc\equiv\sqrt{GM/R}$ & radius $R_{1/2}$\tablenotemark{(a)} & $M_*$\tablenotemark{(b)} & $M_{\rm g}$\tablenotemark{(b)} & $f_{\rm g}$\tablenotemark{(b)} & & clumps\tablenotemark{(c)} &\\
& [$10^{12}\Msun$] & $\frac{M_*}{M_{200}\Omega_b/\Omega_m}$ & [$\kms$] & [$\kpc$] & [$10^{10}\Msun$] & [$10^{10}\Msun$] &  & [$\Msunyr$] &  &\\
\hline
{\it median SINS disk}\tablenotemark{(e)} & - & - & $237$ & $4.6$ & $4.4$ & $1.9$\tablenotemark{(f)} & $30\%$\tablenotemark{(f)} & $65$ & $5.5$ & - \\
\hline
{\it median simulated disk} & $1.5$ & $0.099$ & $238$ & $3.5$ & $2.4$ & $2.1$ & $43\%$ & $27$ & $5$ & $1.31$ \\
\hline
{\it s224} & $1.90$ & $0.129$ & $278$ & $3.3$ & $4.1$ & $2.7$ & $40\%$ & $62$ & $5$ & $1.45$ \\
{\it s263} & $2.18$ & $0.097$ & $289$ & $3.5$ & $3.6$ & $2.1$ & $37\%$ & $38$ & $3$ & $1.51$ \\
{\it s361} & $1.29$ & $0.111$ & $238$ & $4.6$ & $2.4$ & $2.4$ & $50\%$ & $27$ & $0$ & $0.75$ \\
{\it s377} & $1.52$ & $0.084$ & $252$ & $3.3$ & $2.2$ & $1.6$ & $43\%$ & $21$ & $6$ & $1.00$ \\
{\it s383} & $1.73$ & $0.115$ & $253$ & $3.2$ & $3.4$ & $2.1$ & $38\%$ & $29$ & $6$ & $0.86$ \\
{\it s396} & $1.50$ & $0.099$ & $225$ & $3.8$ & $2.5$ & $1.2$ & $33\%$ & $14$ & $5$ & $0.65$ \\
{\it s447} & $1.11$ & $0.089$ & $229$ & $3.5$ & $1.7$ & $1.8$ & $52\%$ & $23$ & $4$ & $1.60$ \\
{\it s466} & $1.39$ & $0.094$ & $238$ & $5.7$ & $2.2$ & $3.0$ & $58\%$ & $30$ & $7$ & $1.41$ \\
{\it s809} & $0.83$ & $0.121$ & $207$ & $2.6$ & $1.7$ & $1.3$ & $43\%$ & $19$ & $5$ & $1.31$ \\
\hline
{\it clump-1} & - & - & - &$0.45$ & $0.022$ & $0.044$ & $67\%$ & $2.7$ & - & - \\
{\it clump-2} & - & - & - &$0.27$ & $0.02$ & $0.04$ & $67\%$ & $2.3$ & - & - \\
{\it clump-3} & - & - & - &$0.22$ & $0.022$ & $0.025$ & $53\%$ & $1.5$ & - & - \\
{\it clump-4} & - & - & - &$0.42$ & $0.027$ & $0.07$ & $72\%$ & $3.7$ & - & - \\
\hline                  
\end{tabular}
\tablecomments{(a) For galaxies: stellar (3D) half-mass radius within $0.2R_{200}$, for clumps: gas half-mass radius inside the region with $\Sigma_g>1000\Msunpc2$ ; (b) For galaxies: inside $3R_{1/2}$, for clumps: inside the region with $\Sigma_g>1000\Msunpc2$ ; (c) Identified as local overdensities in the gas surface density, excluding the galaxy center ; (d) ${\rm\langle SFR\rangle}$ is the mean SFR between $z=3$ and $z=2$ ; (e) Calculated using the nine galaxies from \citet{FoersterSchreiberN_09a} that are defined as 'disks' and have kinematic modeling. Differences to the definitions adopted for the simulations are: $V_c$ is the observed maximum rotation velocity of the $\Ha$-emitting gas; $R_{1/2}$ is calculated from projected $\Ha$ light; all quantities are galaxy-integrated values; the clumps for MD41, BX389 and BX610 are overdensities of rest-frame optical light. Stellar masses and SFRs are estimated with SED modeling, and gas masses are estimated from the SFRs by using the empirical \citet{BoucheN_07a} star-formation relation. Data are taken from \citet{GenzelR_08a,CresciG_09a,FoersterSchreiberN_09a,FoersterSchreiberN_11a}, and are all scaled for a \citet{ChabrierG_03a} initial mass function ; (f) The gas masses are probably lower bounds, since we use the steep \citet{BoucheN_07a} star-formation relation to derive them, as in \citet{FoersterSchreiberN_09a}. If the \citet{KennicuttR_98a} or \citet{GenzelR_10a} relations were used, the gas masses would increase by up to a factor $\approx3$. That would increase the gas fractions to $\approx50\%-60\%$. Indeed, the direct CO observations in \citet{TacconiL_10a} and \citet{DaddiE_10a} indicate $\approx45-65\%$ as the most favourable values.}
\vspace{0.5cm}
\end{table*}

\subsection{Disk and clump properties}
\label{s:disk_and_clump_properties}
\begin{figure*}[tbp]
\centering
\includegraphics[]{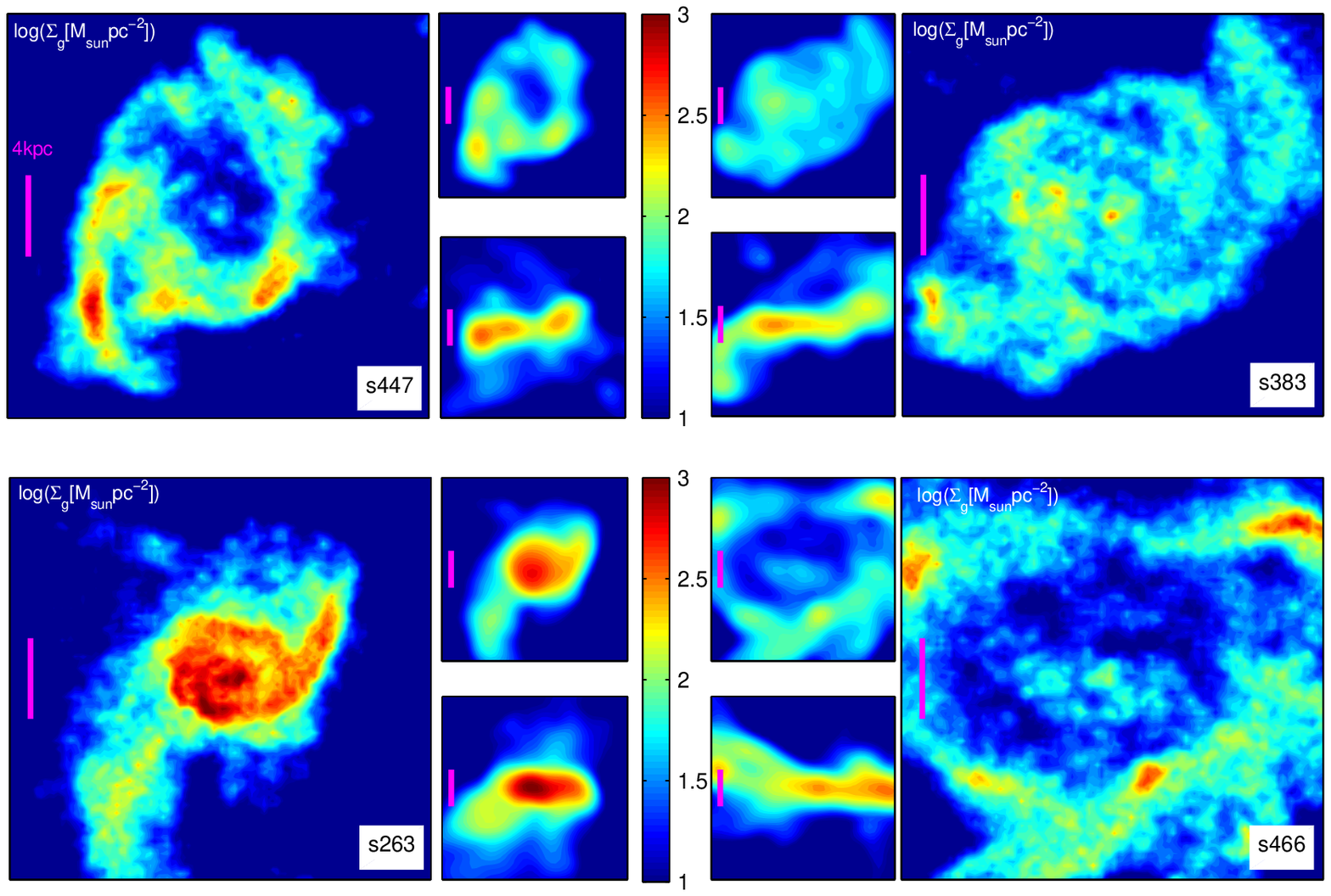}
\caption{Gas surface density maps (in $\Msunpc2$ on a log scale) of four simulated galaxies from our sample. The large panels are maps with resolution compared to the spatial resolution of the simulation. The small panels show the corresponding maps after they have been smoothed with two-dimentional Gaussians with FWHM=$1.5\kpc$, in order to imitate the appearance of the galaxies under the highest available resolution of AO-assisted observations. The small panels show both face-on ({\it top}) and edge-on ({\it bottom}) views. All panels are $20\kpc$ on a side, and all the magenta bars are $4\kpc$ long. It is evident that the qualitative apperance of the simulated galaxies is similar to the clump-clusters and chain galaxies observed at $z\approx2$ \citep{ElmegreenB_05a,ElmegreenB_06a,FoersterSchreiberN_11b}.}
\vspace{0.5cm}
\label{f:SigmaGasMaps}
\end{figure*}

In Figure \ref{f:SigmaGasMaps} we present gas surface density maps of four galaxies from our sample (large panels). The mean gas surface densities inside radii of $\approx4-8\kpc$ have values ranging from $\approx50-300\Msunpc2$. However, the gas surface density is far from constant in the disk, but rather exhibits strong local maxima, to which we hereafter refer as 'clumps'. The perturbed morphologies, and the clumps, are the result of disk instabilities rather than external interactions, as will be discussed in Section \ref{s:clump_evolution}. The typical gas surface densities of clumps are up to ten times higher than the mean disk surface densities, i.e.~$\approx500-3000\Msunpc2$. The clumps tend to be embedded inside transient ring-like features, although not exclusively. The clumpy morphologies of the gas (and hence SFR) are clearly visible also after being smoothed to the resolution of typical good AO-assisted observations (FWHM=$1.5\kpc$), as shown in the small panels, in face-on as well as edge-on views (top and bottom small panels for each galaxy). After this smoothing, the typical contrast ratio (clump-to-disk mean surface densities) is reduced to $\approx3$. The total SFR in all clumps is $\lesssim20\%$ of the total disk SFR. These properties match those of observed $z\approx2$ clumpy disks very well \citep{GenzelR_11a,FoersterSchreiberN_11a}.

\begin{figure*}[tbp]
\centering
\includegraphics[]{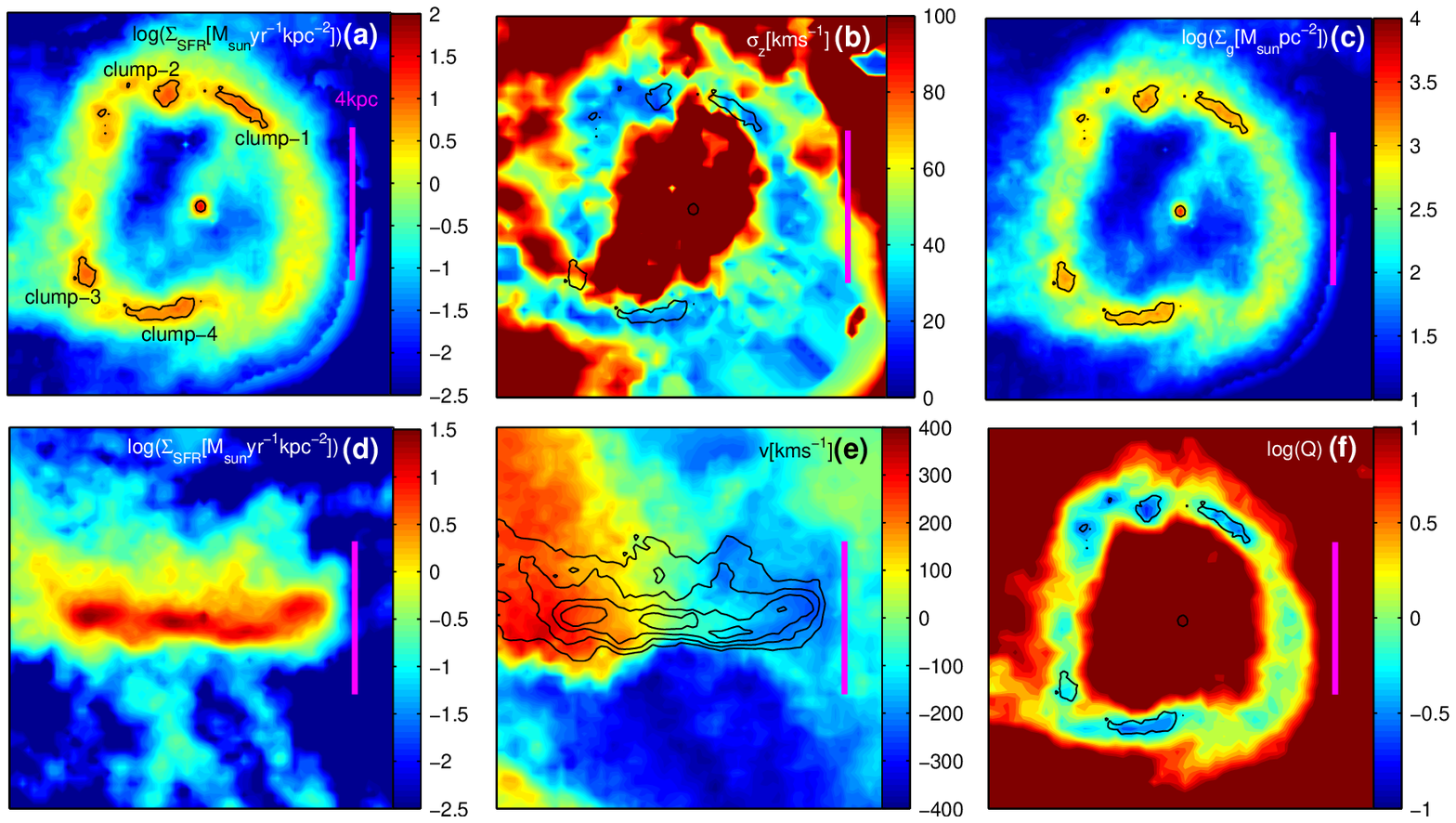}
\caption{Gas properties of the central galaxy {\it s224} at $z=2$. ({\it a}): face-on SFR surface density, ({\it b}): face-on line-of-sight (vertical) velocity dispersion, ({\it c}): face-on gas surface density, ({\it d}): edge-on SFR surface density, ({\it e}): edge-on line-of-sight velocity field overplotted with contours of SFR surface density and ({\it f}): $\log Q$. Each panel is $10\kpc$ on a side. All face-on panels are overplotted with $\Sigma_g=1000\Msunpc2$ contours.}
\vspace{0.5cm}
\label{f:s224_snap284}
\end{figure*}

We now discuss one galaxy from our sample, {\it s224}, in greater detail, as a representative case. The gas is distributed mainly in a ring with a radius of $\approx3\kpc$ (Figure \ref{f:s224_snap284}(c)). The ring, which is transient, formed from spiral features in the disk rather than by an interaction (there are no interactions with stellar mass ratios $<20:1$ since $z\approx3$). Several clumps are embedded in the ring, and appear as SFR overdensities (Figures \ref{f:s224_snap284}(a) and \ref{f:s224_snap284}(d)). The gas vertical velocity dispersion $\sigma_z$ is $\approx20-100\kms$ (Figure \ref{f:s224_snap284}(b)), in agreement with observed values at $z\approx2$. The high density regions and clumps are minima of $\sigma_z$, as denser gas dissipates the random motions more quickly and is less prone to stirring effects such as external accretion \citep{AumerM_10a}. The disk shows regular rotation (Figure \ref{f:s224_snap284}(e)), indicative of its quiet merger history \citep{ShapiroK_08a}.

Hereafter, for {\it s224} we use a threshold of face-on gas surface density $\Sigma_g>1000\Msunpc2$ to define clumps. With this definition, we identify four clumps (Figure \ref{f:s224_snap284}(a)). We calculate their sizes, masses, gas fractions and SFRs inside the region where $\Sigma_g>1000\Msunpc2$, and list them in Table \ref{t:properties}. The clumps have masses of $\approx(5\times10^8-10^9)\Msun$,\footnote{Observed clump masses from the literature tend to be larger, and several clumps with even $\approx10^{10}\Msun$ are known. Several biases can contribute to this apparent discrepancy. First, our simulated disks are less massive than most of the host galaxies of the most massive observed clumps. Therefore, it is the clump-to-disk mass ratio that should be compared. Second, some of the most massive observed clumps have atypical clump-to-disk mass ratios, some of which may be special cases such as clump mergers or galaxy minor mergers. Third, there is no standard for identifying clumps and setting their borders. Such differences may introduce poorly controlled systematics between different studies. The typical clump-to-disk mass ratio reported in \citet{ElmegreenB_05a} is $1\%$, and in \citet{FoersterSchreiberN_11b} it is roughly $2\%$. This is, at face value, $2-3$ times higher than the clump-to-disk mass ratios found in our simulations.} and they are very gas rich ($f_{\rm g}\sim60\%$). The clumps themselves have low ($\approx50\kms\kpc^{-1}$) velocity gradients, not much larger than the galaxy-wide average velocity gradient, which is typically $\approx30\kms\kpc^{-1}$ at a few kiloparsecs from the center.

\subsection{Clump formation and disruption}
\label{s:clump_evolution}
Defining a $Q$ parameter for disk instability \citep{ToomreA_64a} of
\begin{eqnarray}
Q\equiv\frac{\kappa\sqrt{\sigma_g^2+c_s^2}}{\pi G\Sigma_g}
\label{e:Q}
\end{eqnarray}
where $\sigma_g$ is the local gas velocity dispersion, $c_s$ the gas sound speed, and
\begin{eqnarray}
\kappa=\sqrt{3}\left.\sqrt{\frac{GM_{tot}(<R)}{R}}\right/R
\label{e:kappa}
\end{eqnarray}
(e.g.~\citealp{DekelA_09b}), we find that the clumps are local minima with $Q\ll1$ (Figure \ref{f:s224_snap284}(f)). The clumps form in-situ: they are not accreted from outside the disk, but rather form inside the disk via gravitational instability in an environment that has $Q\approx1$. This is consistent with the theoretical framework that is described in the Introduction, and in particular in agreement with other cosmological simulations \citep{AgertzO_09a,CeverinoD_09a,CeverinoD_11a}. The clump masses are $1-2$ orders of magnitude larger than the thermal Jeans mass, but match the 'largest unstable Jeans scale' (the 'Toomre mass') in a turbulent disk
\begin{eqnarray}
&M_{J,\sigma}\approx(\frac{\sigma_g}{V_{\rm rot}})^2M_{\rm disk}\gtrsim\nonumber\\
&(\frac{20}{250})^2\times6\times10^{10}\Msun\approx4\times10^8\Msun
\label{e:M_J}
\end{eqnarray}
\citep{GenzelR_08a,DekelA_09b}, which is driven to high values by the large random motions in the disk.

\begin{figure*}[tbp]
\centering
\includegraphics[]{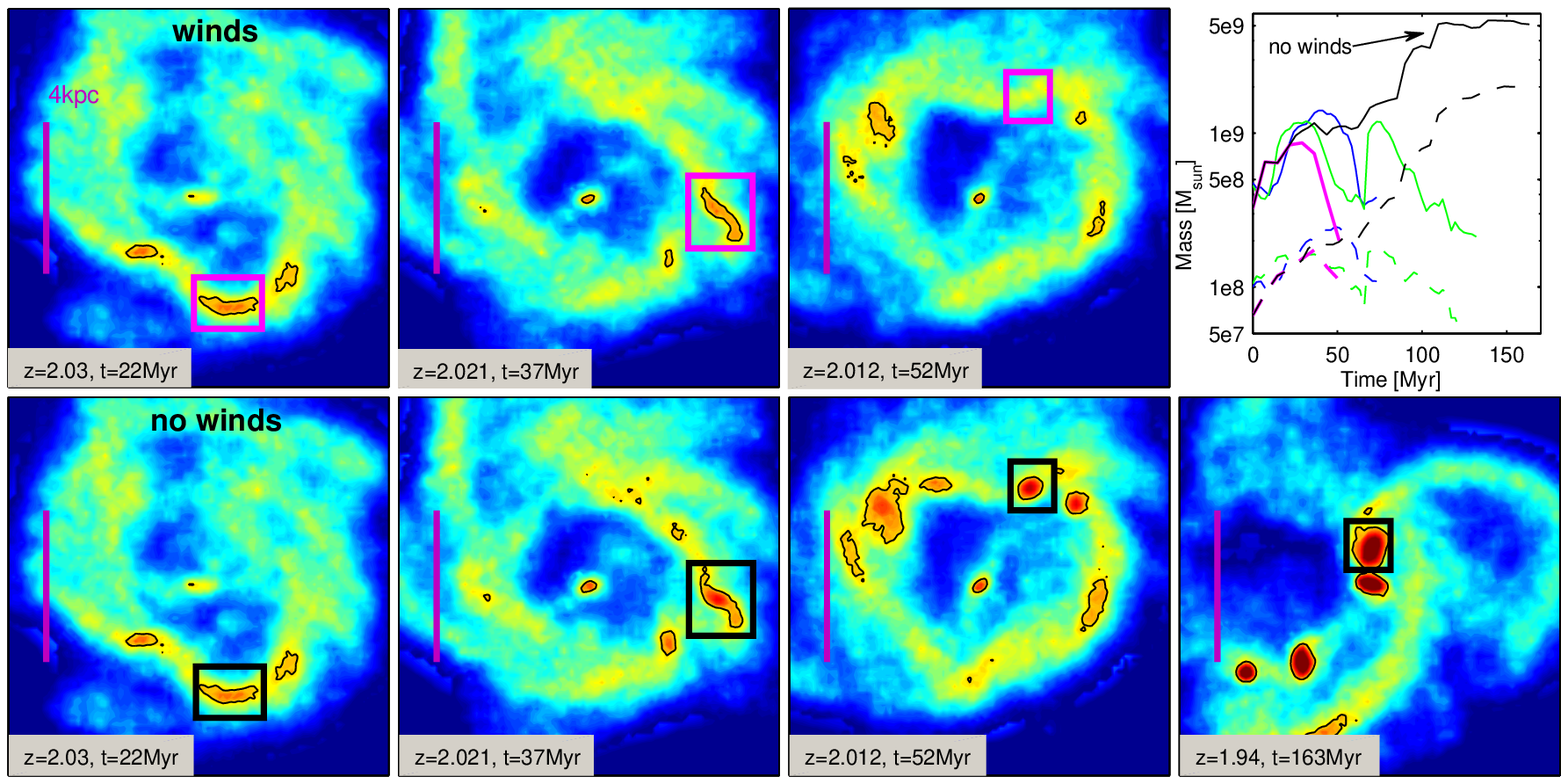}
\caption{A time sequence of gas surface density maps showing the {\it disruption} of a clump in our model ({\it Top}), where $t=0$ (not shown) is the {\it formation} time of the clump. To demonstrate the role of the wind, we turn it off at $z=2.03$ ($t=22\Myr$) and show the alternative evolution of non-disruption, virialisation and migration ({\it Bottom}). The color coding is as in Figure \ref{f:s224_snap284}(c). The upper right-most panel shows the mass of gas ({\it solid}) and young ($<50\Myr$) stars ({\it dashed}) for four clumps as a function of time since their formation. The magenta lines are for the clump highlighted on the {\it Top} and the black for the clump highlighted on the {\it Bottom}. The jump in mass of the green lines at $t\approx60\Myr$ is a result of a merger between two clumps (not shown in other panels). The typical clump lifetime in the presence of winds is $\approx50\Myr$, and the mass of new-formed stars is approximately $10\%$ of the maximum clump gas mass. The mass of new-formed stars internal to the clump decreases following the decrease of the gas mass, as these stars are dispersed out of the clump when the collapse is halted by the return to $Q>1$.}
\vspace{0.5cm}
\label{f:clump_evolution}
\end{figure*}

A notable characteristic of the clumps is their short lifetimes, $\approx50\Myr$, or about half a disk orbital time. The top row in Figure \ref{f:clump_evolution} demonstrates the disruption of a clump in {\it s224} by showing a time series of gas surface density maps. In the upper right-most panel of Figure \ref{f:clump_evolution}, the gas masses of three clumps are shown as a function of time ({\it solid}), demonstrating their rapid formation and disruption. In contrast, the bottom row of Figure \ref{f:clump_evolution} shows the evolution of the same clump in an experiment where we temporarily turn the wind off. The clump does not disrupt, instead it collapses further and virialises, and subsequently migrates to the galaxy center. This demonstrates that the clumps are destroyed by the wind feedback, and do survive (our equation-of-state is sufficiently soft to allow for this), as found in previous simulations, when such feedback is absent (or much weaker than adopted in our sub-grid model).

\begin{figure}[tbp]
\centering
\includegraphics[]{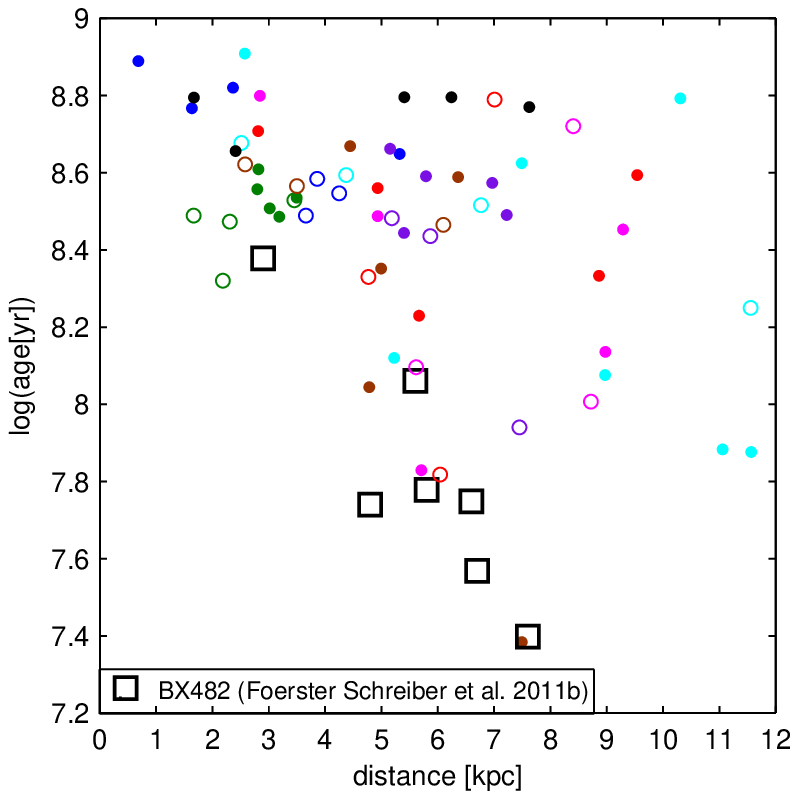}
\caption{Clump stellar age (mass-weighted) versus distance from the galaxy center. Colors correspond to the different galaxies in the same way as in Figure \ref{f:global_relations}, both for $z=2$ ({\it dots}) and $z=2.2$ ({\it circles}). More distant clumps tend to be younger, because of the dominance of 'background' stars in the clumps, and given the global age gradient of the galaxy. A similar relation for the clumps in the observed galaxy BX482 ('raw' measurements, without background subtraction) is shown as well ({\it squares}; \citealp{FoersterSchreiberN_11b}). Given that the observed ages are only relative ages, our simulations seem consistent with the observations of this galaxy. The important conclusion is that the observed trend of clump stellar age with distance (see also \citealp{GenzelR_08a,ElmegreenB_09a}) is not necessarily an indication for clump migration to the center of the galaxy, as has been previously considered.}
\vspace{0.5cm}
\label{f:clump_age_vs_distance}
\end{figure}

The reason for the disruption of clumps by the wind is the following. In our model, the wind mass-loading factor (i.e.~the outflow rate over the SFR) for $10^{10.5}\Msun$ galaxies at $z=2$ is $\eta\approx4$ \citep{OppenheimerB_08a}, and the velocity ranges between $\approx400-700\kms$. These wind parameters agree well with recent observational detections of gas outflowing from $z\approx2$ galaxies \citep{SteidelC_10a} and clumps \citep{GenzelR_11a}. Star-formation in the clumps proceeds on a timescale $T_{SF}\approx300\Myr$ (according to the \citet{KennicuttR_98a} relation), thus winds drive gas out of the clumps on a timescale $\approx T_{SF}/\eta\approx100\Myr$, which is comparable to the disk orbital time. The consequence is that the gas surface density in the clumps decreases faster than the rate at which it is replenished by the instability inside the disk, and so clump regions move from $Q\lesssim1$ to $Q\gtrsim1$ (and thus stop collapsing) within a time that is shorter than the disk orbital time. The short lifetimes we find are consistent with the upper limits of $\approx300-500\Myr$ derived for the stellar populations in the clumps \citep{ElmegreenB_09a,FoersterSchreiberN_11a}. This is because the mean stellar ages of the clumps in our simulations are larger than the lifetimes of the clumps as star-forming overdensities (see Figure \ref{f:clump_age_vs_distance} and accompanying discussion below).

Given the masses and sizes of clumps (Table \ref{t:properties}), their circular velocities equal $\sqrt{GM/R}\approx80-100\kms$. This is significantly larger than the velocity dispersion of the gas within them or their rotational velocities (in particular if the galaxy-wide average velocity gradient is subtracted as 'background'). Typically the clumps have $v_{\rm rot}^2+2\sigma_g^2\approx0.6(GM/R)$, or $|\frac{U}{K}|\gtrsim3$, where $U$ is their gravitational potential energy and $K$ the internal kinetic energy. Thus, they are neither supported by pressure nor by rotation, and in fact they are not virialised. This is due to continuous feedback on a timescale shorter than the dynamical time, preventing virialisation before disruption.

The clumps consist two types of stellar populations. First, `background stars', which are part of the stellar disk at the locations where clumps form but are not necessarily bound to them, with an age distribution similar to that of the stellar disk as a whole. A dynamically cold stellar component also takes part in the instability, in addition to the gas, such that the local density of old `background stars' is increased in some clumps compared to the clump surroundings. The `background stars' comprise most of the stellar mass of the clumps, as the clumps typically convert just about $10\%$ of their peak gas mass into stars throughout their lifetime. Second are `clump stars', which form over the lifetime of a clump, and are almost all bound to their clumps, constituting typically up to $\approx20\%$ of their {\it stellar} masses.

As a result, the mean age of the stellar populations inside clumps follows the global spatial age distribution of the 'background stars' of the disk, with an additional small contribution of the 'clump stars', which makes the clumps typically be local age minima. We find that the stellar disks in our simulations have a negative age (and metallicity) gradient, which will be discussed in detail in future work. Therefore, the mean ages of clumps also present a spatial gradient, such that clumps with older stellar populations are closer to the galaxy center. This is shown in Figure \ref{f:clump_age_vs_distance}. The trend we find is qualitatively consistent with observations, in spite of the fact that no individual clump actually migrates to the galaxy center as its stellar populations become older. A quantitative comparison to observations cannot be performed yet, as measuring clump ages requires high-resolution, deep data taken with various instruments. For example, \citet{FoersterSchreiberN_11b} was able to measure clump ages for only one $z\approx2$ galaxy from their sample, BX482, which are shown in Figure \ref{f:clump_age_vs_distance} ({\it squares}). These measurements rely on $\Ha$ equivalent width, and they are very sensitive to the assumptions made regarding the star-formation history. Therefore, they should be taken only as relative ages, not absolute ones \citep{FoersterSchreiberN_11b}. A linear fit in the log-log plane to the clump age-distance relation from our simulations yields a slope of $-0.57\pm0.14$. The corresponding slope for the BX482 points shown in Figure \ref{f:clump_age_vs_distance} is $-2.06\pm0.63$. Future comparisons of this correlation with larger galaxy samples, and with an improved treatment of possible biases both on the theoretical and observational sides, may be able to distinguish the slope we find from the $-1.05$ slope found by \citet{CeverinoD_11a} for the case of migrating clumps. To conclude, the mere existence of an age gradient should not necessarily be interpreted as supporting the clump migration scenario (see also \citet{FoersterSchreiberN_11b} for additional alternative interpretations).

\section{Direct comparison to SINFONI observations}
\label{s:comparison}
\begin{figure*}[tbp]
\centering
\includegraphics[]{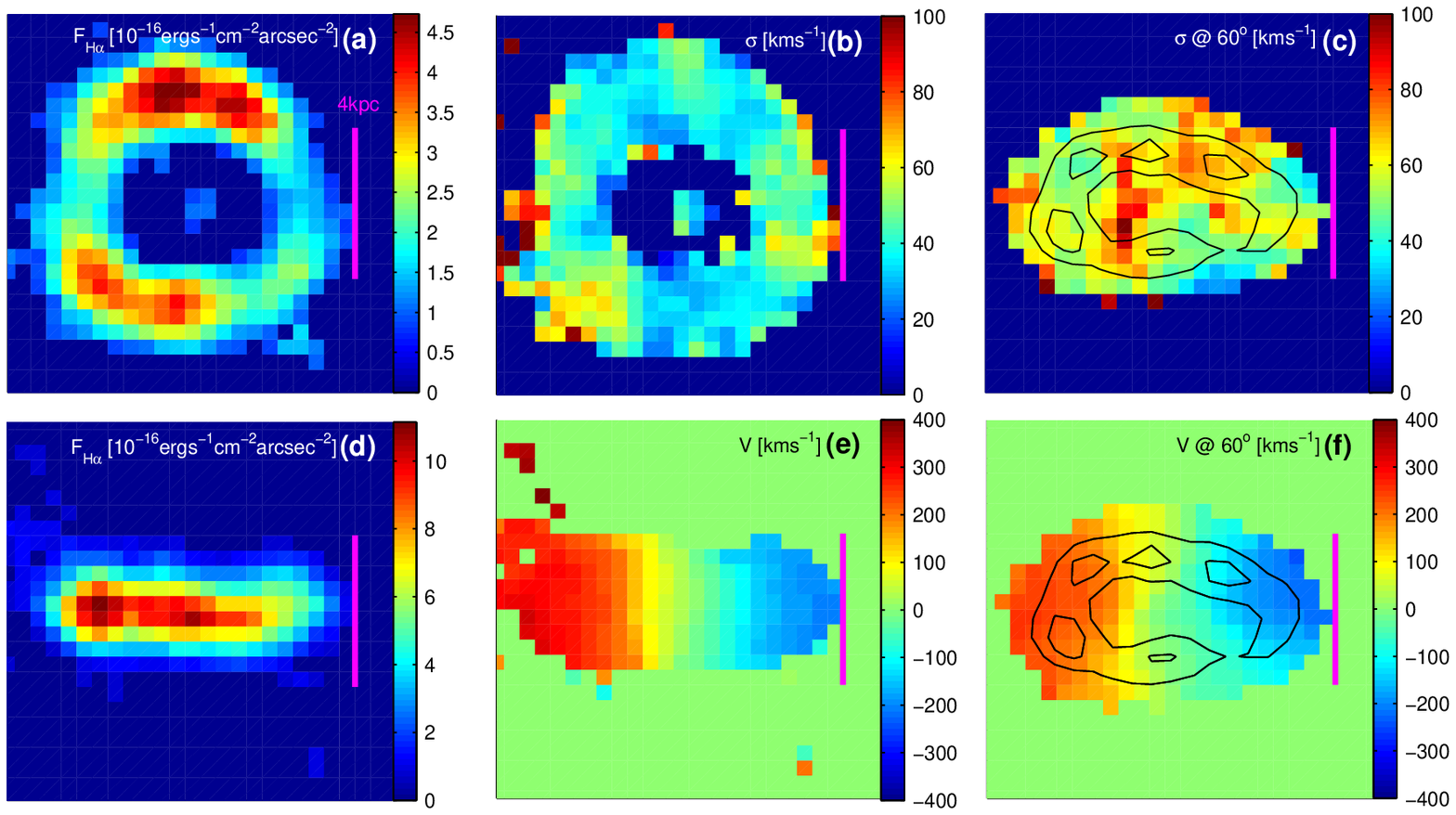}
\caption{Mock $\Ha$ line observations of the snapshot shown in Figure \ref{f:s224_snap284}, as would be obtained with SINFONI. Face-on ({\it a}) and edge-on ({\it d}) line intensity, velocity field shown edge-on ({\it e}) and at $60^{\circ}$ inclination ({\it f}), and velocity dispersion shown face-on ({\it b}) and at $60^{\circ}$ inclination ({\it c}). Contours of surface density are overplotted in panels (c) and (f). See Section \ref{s:comparison} for discussion.}
\vspace{0.5cm}
\label{f:mock_SINS}
\end{figure*}

Figure \ref{f:mock_SINS} shows mock SINFONI maps for {\it s224} of $\Ha$ intensity, velocity and velocity dispersion. They were obtained by converting SFR to $\Ha$ luminosity using $L_{\Ha}[\ergs]=1.26\times10^{41}\times {\rm SFR}[\Msunyr]$ \citep{KennicuttR_98a}, `placing' {\it s224} at $z=2.2$, convolving it with a Gaussian to imitate a resolution of ${\rm FWHM}=0.17''$, and pasting it into a real SINFONI datacube with pixel scale $0.05''$ that was obtained from the observations of a galaxy from the SINS survey, at a redshift where no real galaxy exists. This results in realistic resolution and noise properties that correspond to a representative total integration time ($\approx6$ hours) for our high-resolution SINFONI data sets \citep{FoersterSchreiberN_09a,GenzelR_11a}.

The clumpiness, smooth velocity field and relatively flat velocity dispersion map (outside the center), which are the characteristics of real SINS clumpy disks, are all well reproduced when 'observing' our simulations. The most significant difference to the non-degraded images in Figure \ref{f:s224_snap284} (the left and middle columns can be directly compared in Figures \ref{f:mock_SINS} and \ref{f:s224_snap284}) is in the velocity dispersion map. The 'beam smearing' increases the apparent velocity dispersion where there are velocity gradients (this is the reason for the diagonal feature of high dispersion in the inclined image in Figure \ref{f:mock_SINS}(c)), and the clumps are no longer seen as clear minima. As a result, the small but present velocity gradients across the clumps get smoothed out. Thus, because of the non-virialised state of the clumps, and their being minima in velocity dispersion, they do not show strong features in observations at the currently available resolution of $\approx1-2\kpc$, even if their masses are dynamically significant. This prediction is consistent with the observations of \citet{GenzelR_11a} (but see \citet{CeverinoD_11a} for alternative explanations).

\section{Summary and discussion}
\label{s:discussion}
In this paper we use 'zoom-in' cosmological hydrodynamical simulations that include a phenomenological model for efficient star-formation feedback in the form of fast momentum-driven winds with high mass-loading factors, to investigate the formation of star-forming disk galaxies at $z\approx2$. Our wind model reproduces observed low baryon conversion efficiencies and high gas fractions, and we also find a good kinematical and morphological agreement with observations. In addition, we present an important numerical result, namely that in the presence of such winds, the giant clumps in $z\approx2$ disks, which form in-situ via gravitational instability, may be short-lived and unvirialised. This is because they lose mass to galactic outflows at a rate that is higher than the rate of their dynamical evolution, such that their collapse is halted in less than a disk orbital time after their formation, and before they reach virial equilibrium. We show that this scenario is plausibly consistent with most of the available observations of clumpy disks at $z\approx2$: their global properties (Figure \ref{f:global_relations}), their clumpiness (Figures \ref{f:SigmaGasMaps} and \ref{f:s224_snap284}), their kinematics and appearance in SINFONI observations (Figures \ref{f:s224_snap284} and \ref{f:mock_SINS}), and the gradient they show between clump stellar age and clump distance from the galaxy center (Figure \ref{f:clump_age_vs_distance}).

Our results are in contrast to previous models that do not include fast winds with high mass-loading factors. We demonstrate that if we temporarily turn off the winds, we reproduce previous results where clumps are long-lived and migrate to the galaxy centers. Therefore we are able to directly show that the winds are the critical component of our model that makes the clumps short-lived. However, if the wind is drastically reduced starting from the initial conditions of the simulations at high redshift, gas-rich unstable disks do not develop in the first place. Instead, the galaxies become gas-poor, very massive (compared to the halos inside which they form, i.e.~with very high baryon conversion efficiencies), and they develop very peaked rotation curves (see also \citealp{JoungM_09a}).

It should be stressed that our model of the winds is phenomenological, and our results are based on the working hypothesis of the existence of powerful superwinds with some given, a priori properties. Our hypothesis still awaits further verification. For example, our feedback recipe assumes that gas can be blown out of the clumps and subsequently out of the galaxy at an imposed rate proportional to the star-formation rate. It remains to be better understood how this occurs on sub-clump scales based on improved physical models (e.g.~\citealp{HopkinsP_11a}). Also, the momentum we inject into the wind per unit stellar mass formed is at the high side of the theoretical expectations. While winds are observed to be ubiquitous in high-redshift galaxies, and in particular they have been recently observed from individual giant clumps, the measurements of their crucial parameters, in particular the mass-loading factors, is still very uncertain.

It has been suggested in the recent literature that the migration of giant clumps to the centers of high-redshift galaxies may serve as an avenue for the formation of galaxy bulges, in addition to the more traditional avenue of galaxy mergers (e.g.~\citealp{SteinmetzM_02a}). The alternative scenario we present in this paper, namely that giant clumps do not migrate to their galaxy centers, does not necessarily mean that bulges do not form secularly in high-redshift galaxy disks. The instabilities and irregularities we find in our simulated disks do induce internal torques and angular momentum loss, such that inflows to smaller radii inside the disks do occur. Nevertheless, by comparing clump evolution with and without winds, we demonstrate that secular bulge growth occurs much more slowly in the case where winds disrupt clumps. We reserve detailed studies of secular bulge formation under conditions of strong feedback and giant clump destruction, as well as implications for the global cosmic rate of bulge formation, to future work.

\acknowledgements
We thank Jerry Ostriker for useful discussions. The high performance computations were performed on the HLRB-II system provided by LRZ and on the SGI-Altix 3700 Bx2 (University Observatory Munich), partly funded by the Cluster of Excellence: "Origin and Structure of the Universe". We thank the DFG for support via German-Israeli Project Cooperation grant STE1869/1-1.GE625/15-1. SG acknowledges the PhD fellowship of the International Max Planck Research School in Astrophysics. N.M.F.S. acknowledges support by the Minerva program of the MPG.

\end{document}